\documentclass[journal,twoside]{IEEEtran}
\IEEEoverridecommandlockouts
\usepackage{cite}

\ifCLASSINFOpdf
  \usepackage[pdftex]{graphicx}
  \graphicspath{}
  \DeclareGraphicsExtensions{.pdf,.jpeg,.png}
\else
  \usepackage[dvips]{graphicx}
  \graphicspath{}
  \DeclareGraphicsExtensions{.eps}
\fi
\usepackage[cmex10]{amsmath}
\usepackage{amssymb}
\ifCLASSOPTIONcompsoc
  \usepackage[caption=false,font=normalsize,labelfont=sf,textfont=sf]{subfig}
\else
  \usepackage[caption=false,font=footnotesize]{subfig}
\fi
\usepackage{abbrevs}
\usepackage{tikz}

\usepackage{mathrsfs}
\usepackage{multirow}

\newtheorem{theorem}{Theorem}

\begin{document}
\tikzstyle{state}=[shape=circle,draw=blue!90,fill=blue!10,line width=1pt]
%
\title{Bayesian Smoothing for the Extended Object Random Matrix Model}

\author{Karl~Granstr\"om,~\IEEEmembership{Member,~IEEE},
        Jakob~Bramst{\aa}ng
\thanks{Karl Granstr{\"o}m is with the Department of Electrical Engineering, Chalmers University of Technology, Gothenburg, Sweden. E-mail: \texttt{karl.granstrom@chalmers.se}.

Jakob Bramst{\aa}ng did his part of this work as a Master's Student at the Department of Electrical Engineering, Chalmers University of Technology, Gothenburg, Sweden. He is currently with Knightec AB, Stockholm, Sweden. E-mail: \texttt{jakob.bramstang@knightec.se}.}%
}

%


\maketitle

\begin{abstract}
The random matrix model is popular in extended object tracking, due to its relative simplicity and versatility. In this model, the extended object state consists of a kinematic vector for the position and motion parameters (velocity, etc), and an extent matrix. Two versions of the model can be found in literature, one where the state density is modelled by a conditional density, and one where the state density is modelled by a factorized density. In this paper, we present closed form Bayesian smoothing expression for both the conditional and the factorised model. In a simulation study, we compare the performance of different versions of the smoother.
\end{abstract}


%
\IEEEpeerreviewmaketitle

\section{Introduction}


Multiple Object Tracking (\mot) denotes the process of successively determining the number and states of multiple dynamic objects based on noisy sensor measurements. Tracking is a key technology for  many technical applications in areas such as robotics, surveillance, autonomous driving, automation, medicine, and sensor networks. Extended object tracking is defined  as \mot where each object generates multiple measurements per time step and the measurements are spatially structured on the object, see \cite{GranstromBR:2017}. 

Extended object tracking is applicable in many different scenarios, e.g., environment perception for autonomous vehicles using camera, lidar and automotive radar. The multiple measurements per object and time step create a possiblity to estimate the object extent, in addition to the position and the kinematic properties such as velocity and heading. This estimation requires an object state space model, including modelling of the object dynamics and the measurement process. Extended obejct models include the Random Matrix model \cite{Koch:2008,FeldmannFK:2011}, the Random Hypersurface model \cite{BaumH:2014}, and Gaussian Process models \cite{WahlstromO:2015}. A comprehensive overview of extended object tracking can be found in \cite{GranstromBR:2017}. 

In this paper we focus on the Random Matrix model, also known as the Gaussian inverse Wishart (\giw) model. The random matrix model was originally proposed by Koch \cite{Koch:2008}, and is an example of a spatial model. In this model the shape of the object is assumed to be elliptic. The ellipse shape is simple but still versatile, and the random matrix model has been integrated into many different multiple extended object tracking frameworks \cite{WienekeK:2010,WienekeD:2011,WienekeK:2012,GranstromO:2012a,LundquistGO:2013,BeardRGVVS:2016,GranstromFS:2016fusion,SchusterRW:JAIF,SchusterRW:2015,VivoneB:2016}. Indeed, the random matrix model is applicable in many real scenarios, e.g., pedestrian tracking using video \cite{WienekeK:2010,WienekeD:2011,WienekeK:2012} or lidar \cite{GranstromO:2012a} and tracking of boats and ships using marine radar \cite{GranstromNBLS:2014,GranstromNBLS:2015_TGARS,VivoneB:2016,VivoneBGW:2016,VivoneBGNC:2015_ConvMeas,VivoneBGNC:2017,SchusterRW:2015}.

The focus of this paper is on Bayesian smoothing for the random matrix model. A preliminary version of this work was presented in \cite{Bramstang:2018}. This paper is a significant extension of \cite{Bramstang:2018}, and presents the following contributions:
\begin{itemize}
	\item Closed form smoothing expressions for the conditional \giw model from \cite{Koch:2008}.
	\item Closed from smoothing expressions for the factorized \giw model from \cite{FeldmannFK:2011}. 
	\item A simulation study that compares the derived smoothers to both prediction and filtering.
\end{itemize}
As a minor contribution, a closed form expression for the random matrix prediction from \cite{GranstromO:2014} is presented.

The rest of the paper is organized as follows. A problem formulation is given in the next section. In Section~\ref{sec:randomMatrixReview} a review of the random matrix model is given. Smoothing for the conditional \giw model is presented in Section~\ref{sec:condSmoothing}; smoothing for the factorised \giw model is presented in Section~\ref{sec:factSmoothing}. Results from a simulation study are presented in Section~\ref{sec:simulationStudy}. Concluding remarks are given in Section~\ref{sec:Conclusions}.

\begin{table}
	\caption{Notation}
	\label{tab:notation}
\vspace{-5mm}
\rule[0pt]{\columnwidth}{1pt}
	\begin{itemize}
		\item $\mathbb{R}^{n}$: space of vectors of dimension $n$
		\item $\mathbb{I}^{n\times n}$: space of non-singular $n\times n$ matrices
		\item $\mathbb{S}_{+}^{d}$: space of positive semi-definite $d\times d$ matrices
		\item $\mathbb{S}_{++}^{d}$: space of positive definite $d\times d$ matrices
		\item $\Id$: unit matrix of size $d\times d$
		\item $\mathbf{0}_{m\times n}$: all-zero $m\times n$ matrix
		\item $\otimes$: Kronecker product
		\item $|\cdot |$: set cardinality
		\item $\diag{\cdot}$: diagonal matrix
		\item $\mathbb{E}[\cdot]$: expected value
		\item $ \Npdfbig{\sx}{m}{P}$: Gaussian pdf for random vector $\sx\in\mathbb{R}^{n_x}$ with mean vector $m\in\mathbb{R}^{n_x}$ and covariance matrix $P\in\mathbb{S}_{+}^{n_x}$
		\item $\IWishpdf{\ext}{v}{V}$: inverse Wishart pdf for random matrix $\ext\in\mathbb{S}_{++}^{d}$ with degrees of freedom $v>2d$ and parameter matrix $V\in\mathbb{S}_{++}^{d}$, see, e.g., \cite[Def. 3.4.1]{GuptaN:2000}
		\item $\Wishpdf{\ext}{v}{V}$: Wishart pdf for random matrix $\ext\in\mathbb{S}_{++}^{d}$ with degrees of freedom $v\geq d$ and parameter matrix $V\in\mathbb{S}_{++}^{d}$, see, e.g., \cite[Def. 3.2.1]{GuptaN:2000}
		\item $\mathcal{GB}_{d}^{II}\left(\ext \ ; \ a,\ b,\ \Omega,\ \Psi\right)$: Generalized matrix variate beta type II pdf for random matrix $\ext\in\mathbb{S}_{++}^{d}$ with degrees of freedom $a>\frac{d-1}{2}$, $b>\frac{d-1}{2}$, parameter matrix $\Psi\in\mathbb{S}_{+}^{d}$, and parameter matrix $\Omega$ such that $(\Omega-\Psi)\in\mathbb{S}_{++}^{d}$, see, e.g., \cite[Def. 5.2.4]{GuptaN:2000}
		
	\end{itemize}
\rule[0pt]{\columnwidth}{1pt}
\end{table}


\section{Problem formulation}

Let $\xi_{k} $ denote the extended object state at time $k$,  let $\setZ_{k}$ denote the set of measurements at time step $k$, and let $\setZ_{1:k}$ denote the sets of measurements from time $1$ up to, and including, time $k$. Bayesian extended object filtering builds upon two steps, the Chapman-Kolmogorov prediction
\begin{align}
	p(\xi_{k+1} | \setZ_{1:k}) & = \int p(\xi_{k+1}|\xi_{k}) p(\xi_{k} | \setZ_{1:k}) \diff \xi_{k}
	\label{eq:SmoothingForwards}
\end{align}
where $p(\xi_{k+1}|\xi_{k})$ is the transition density, and the Bayes update
\begin{align}
	p(\xi_{k+1} | \setZ_{1:k+1}) & = \frac{p(\setZ_{k+1} | \xi_{k+1}) p(\xi_{k+1} | \setZ_{1:k})}{ \int p(\setZ_{k+1} | \xi_{k+1}) p(\xi_{k+1} | \setZ_{1:k}) \diff \xi_{k+1}}
\end{align}
where $p(\setZ_{1:k+1} | \xi_{k+1})$ is the measurement likelihood. The focus of this paper is on Bayesian extended object smoothing, 
\begin{align}
	p(\xi_{k} | \setZ_{1:K}) & = p(\xi_{k} | \setZ_{1:k}) \int \frac{p(\xi_{k+1}|\xi_{k}) p(\xi_{k+1}|\setZ_{1:K})}{p(\xi_{k+1}|\setZ_{1:k})} \diff \xi_{k+1},
	\label{eq:SmoothingBackwards}
\end{align}
where $K$ is the final time step. For the random matrix model, the Chapman-Kolmogorov prediction and Bayes update have been covered extensively in previous litterature, see, e.g., \cite{Koch:2008,FeldmannFK:2011,GranstromO:2014,LanRL:2012} for the prediction, and, e.g., \cite{Koch:2008,FeldmannFK:2011,Orguner:2012,ArdeshiriOG:2015,SaritasO:2018,LanRL:2012} for the update. In this paper, we focus on Bayesian extended object smoothing. In previous literature, smoothing is only discussed briefly in \cite[Sec. 3.F]{Koch:2008}, and complete details are not given. 

Bayesian filtering and smoothing for the random matrix model is an example of \emph{assumed density filtering}: the functional form of the state density is to be preserved in the prediction and the update. It is therefore necessary that Bayesian smoothing also preserves the functional form of the extended object state density. Two different assumed state densities can be found in the literature: the conditional Gaussian inverse Wishart \cite{Koch:2008}, and the factorized Gaussian inverse Wishart \cite{FeldmannFK:2011}.  

The problem considered in this paper is to use the Bayesian smoothing equation \eqref{eq:SmoothingBackwards} to compute the smoothing \giw parameters for both the conditional model and the factorized model.

\section{Review of random matrix model}
\label{sec:randomMatrixReview}

In this section we give a brief review of the random matrix model; a longer review can be found in \cite[Sec. 3.A]{GranstromBR:2017}.

In the random matrix model \cite{Koch:2008,FeldmannFK:2011}, the extended object state is a tuple $\xi_{k} = \left(\sx_{k},\ext_{k}\right) \in \mathbb{R}^{n_x} \times \mathbb{S}_{++}^{d}$.  The vector $\sx_{k} \in \mathbb{R}^{n_x}$ represents the object's position and its motion properties, such as velocity, acceleration, and turn-rate. The matrix $\ext_{k} \in \mathbb{S}_{++}^{d}$ represents the object's extent, where $d$ is the dimension of the object; $d=2$ for tracking with 2D position and $d=3$ for tracking with 3D position. The matrix $\ext_{k}$ is modelled as being symmetric and positive definite, which means that the object shape is approximated by an ellipse. 

In the literature, there are two alternative models for the extended object state density, the conditional and the factorised. In the conditional model, first presented in \cite{Koch:2008}, the following state density is used,
\begin{subequations}
\begin{align}
	p(\xi_{k} | \setZ_{1:\ell}) = & p\left(\sx_{k}|\ext_{k},\setZ_{1:\ell}\right)p\left(\ext_{k}|\setZ_{1:\ell}\right) \label{eq:conditionalDensityGeneral} \\
= & \Npdfbig{\sx_{k}}{m_{k|\ell}}{P_{k|\ell}\otimes\ext_{k}} \nonumber \\
& \times \IWishpdf{\ext_{k}}{v_{k|\ell}}{V_{k|\ell}}, \label{eq:conditionalGIW}%
\end{align}%
\end{subequations}%
where $m_{k|\ell}\in\mathbb{R}^{n_x}$, $P_{k|\ell}\in\mathbb{S}_{+}^{s}$, $v_{k|\ell}>2d$, $V_{k|\ell}\in\mathbb{S}_{++}^{d}$, and $s=\frac{n_x}{d}$. In this model, the random vector $\sx$ consists of a $d$-dimensional spatial component (the position) and its derivatives (velocity, acceleration, etc.), see \cite[Sec. 3]{Koch:2008}. Thus,  $s-1=\frac{n_x}{d}-1$ describes up to which derivative the kinematics are described, see \cite[Sec. 3]{Koch:2008}.

In the factorised model, first presented in \cite{FeldmannFK:2011}, the following state density is used,
\begin{subequations}
\begin{align}
	p(\xi_{k} | \setZ_{1:\ell}) = & p\left(\sx_{k}|\setZ_{1:\ell}\right)p\left(\ext_{k}|\setZ_{1:\ell}\right) \\
= & \Npdfbig{\sx_{k}}{m_{k|\ell}}{P_{k|\ell}} \IWishpdf{\ext_{k}}{v_{k|\ell}}{V_{k|\ell}},
\end{align}%
\label{eq:factorizedGIW}%
\end{subequations}%
where $m_{k|\ell}\in\mathbb{R}^{n_x}$, $P_{k|\ell}\in\mathbb{S}_{+}^{n_x}$, $v_{k|\ell}>2d$, and $V_{k|\ell}\in\mathbb{S}_{++}^{d}$. In this model, the random vector $\sx$ consists of a $d$-dimensional spatial component (the position) and additional motion parameters; note that, in contrast to the conditional model, here the motion parameters are not restricted to being derivatives of the spatial component, and non-linear dynamics can be modelled, see further in \cite{FeldmannFK:2011}.

%
%
%

The random matrix transition density can expressed as
\begin{subequations}
\begin{align}
	p(\xi_{k+1}|\xi_{k}) = & p \left(\sx_{k+1},\ext_{k+1}|\sx_{k},\ext_{k}\right) \\
	= & p \left(\sx_{k+1}|\ext_{k+1},\sx_{k},\ext_{k}\right) p \left(\ext_{k+1}|\sx_{k},\ext_{k}\right) \\
	= & p \left(\sx_{k+1}|\ext_{k+1},\sx_{k}\right) p \left(\ext_{k+1}|\sx_{k},\ext_{k}\right) \label{eq:randomMatrixTransitionDensity}
\end{align}
\end{subequations}
where the last equality follows from a Markov assumption, see \cite{Koch:2008}. The random matrix measurement likelihood can be expressed on a general form as
\begin{align}
	p(\setZ_{k} | \xi_{k}) \propto \prod_{\sz\in\setZ_{k}} p(\sz | \sx_k,\ext_k) 
\end{align}
Note that the modelling of the extended object measurement set cardinality is outside the scope of this work, see \cite[Sec. 2.C]{GranstromBR:2017} for an overview of different models for the number of measurements. 


\section{Conditional model smoothing}
\label{sec:condSmoothing}

In the conditional model, we have conditional Gaussian inverse Wishart densities, cf. \eqref{eq:conditionalGIW},
and under assumed density filtering we seek a smoothed density of the same form, i.e.,
\begin{subequations}
\begin{align}
	p(\xi_{k} | \setZ_{1:K}) = & p(\sx_{k}|\ext_{k},\setZ_{1:K})p(\ext_{k} | \setZ_{1:K}) \\
	= & \Npdfbig{\sx_{k}}{m_{k|K}}{P_{k|K}\otimes\ext_{k}} \nonumber \\
& \times \IWishpdf{\ext_{k}}{v_{k|K}}{V_{k|K}}.
\end{align}
\label{eq:condSmoothedGIW}
\end{subequations}

\subsection{Assumptions and modelling}
The following assumptions are made for the conditional \giw model, see \cite[Sec. 2]{Koch:2008}.
\begin{assumption}\label{ass:ExtentTransitionDensity}
	The time evolution of the extent state is assumed independent of the kinematic state,
	\begin{align}
		p\left(\ext_{k+1}|\sx_{k},\ext_{k}\right) = p\left(\ext_{k+1}|\ext_{k}\right).
	\end{align}
	\hfill$\square$
\end{assumption}
\begin{assumption}\label{ass:SlowlyChangingExtent}
	The extent changes slowly with time, $\ext_{k+1} \approx \ext_{k}$, such that for the kinematic state, conditioned on the extent state, the following holds,
	\begin{align}
		p(\sx_{k} | \ext_{k}) & \approx p(\sx_{k} | \ext_{k+1}), \\
		p(\sx_{k+1} | \ext_{k+1}) & \approx p(\sx_{k+1} | \ext_{k}), \\
		p\left(\sx_{k+1}|\ext_{k+1},\sx_{k}\right) & \approx p\left(\sx_{k+1}|\ext_{k},\sx_{k}\right).
	\end{align}
	\hfill$\square$
\end{assumption}
The validity of Assumptions~\ref{ass:ExtentTransitionDensity} and \ref{ass:SlowlyChangingExtent} is discussed in \cite{Koch:2008}.

In the conditional random matrix model, the transition density \eqref{eq:randomMatrixTransitionDensity} is Gaussian-Wishart, see \cite[Sec. 3.A/B]{Koch:2008},
\begin{subequations}
\begin{align}
	p(\xi_{k+1}|\xi_{k}) 
	\approx & p\left(\sx_{k+1}|\ext_{k+1},\sx_{k}\right)p\left(\ext_{k+1}|\ext_{k}\right) \label{eq:condTransitionDensityGeneral} \\
= & \Npdfbig{\sx_{k+1}}{(F_k\otimes \mathbf{I}_{d})\sx_{k}}{D_k \otimes \ext_{k+1}} \\
& \times \Wishpdf{\ext_{k+1}}{n_{k}}{\frac{\ext_{k}}{n_{k}}} \nonumber
\end{align}
where the $s\times s$ matrix $F_k$ is the motion model, the $s\times s$ matrix $D_k$ is the process noise, and the degrees of freedom $n_k \geq d$ govern the uncertainty of the time evolution of the extent. This transition density was generalised by \cite{LanRL:2012} by introducing a $d\times d$ parameter matrix $A$ for the extent transition,
\begin{align}
	p(\xi_{k+1}|\xi_{k}) \approx & \Npdfbig{\sx_{k+1}}{(F_k\otimes \mathbf{I}_{d})\sx_{k}}{D_k \otimes \ext_{k+1}} \label{eq:condTransitionDensity} \\
& \times \Wishpdf{\ext_{k+1}}{n_{k}}{\frac{A\ext_{k} A^{\tp}}{n_{k}}} \nonumber
\end{align}
\end{subequations}
In the remainder of the paper, we consider this generalised transition density. The measurement model is \cite[Sec. 3.D]{Koch:2008}
\begin{align}
	p(\sz | \sx_k,\ext_k) = \Npdfbig{\sz}{\left(H_k \otimes \mathbf{I}_d \right)\sx_{k}}{\ext_{k}},
	\label{eq:condMeasurementModel}
\end{align} 
where the $1 \times s$ matrix $H_k$ is the measurement model. 

\begin{table}[t]
\caption{Conditional model: prediction}
\label{tab:condPrediction}
\vspace{-5mm}
\rule[0pt]{\columnwidth}{1pt}
\begin{align*}
	m_{k+1|k} &= \left(F_{k}\otimes\Id\right) m_{k|k} \\
	P_{k+1|k} &= F_{k} P_{k|k} F_{k}^{\tp} + D \\
	v_{k+1|k} &= d+1 + \left(1 + \frac{v_{k|k}-2d-2}{n}\right)^{-1}(v_{k|k}-d-1) \\
	V_{k+1|k} &= \left(1 + \frac{v_{k|k}-d-1}{n-d-1}\right)^{-1} A V_{k|k} A^{\tp}
\end{align*}
\rule[0pt]{\columnwidth}{1pt}
\end{table}
\begin{table}[t]
\caption{Conditional model: update}
\label{tab:condUpdate}
\vspace{-5mm}
\rule[0pt]{\columnwidth}{1pt}
\begin{align*}
\begin{array}{rcl}
m_{k|k} &= & m_{k|k-1} + (K \otimes \mathbf{I}_{d}) \varepsilon \\
P_{k|k} &= & P_{k|k-1} - K S K^{\tp} \\
v_{k|k} &= & v_{k|k-1} + |\setZ_{k}| \\
V_{k|k} &= & V_{k|k-1} + N + Z \\
\varepsilon &= &  \bar{\sz} - (H \otimes \mathbf{I}_{d}) m_{k|k-1}^{}\\
\bar{\sz}  &= &  \frac{1}{|\setZ_{k}|}\sum_{\sz \in \setZ_{k}}{\sz} \\
Z &= & \sum_{\sz\in \setZ_{k}} \left(\sz - \bar{\sz}\right)\left(\sz - \bar{\sz}\right)^{\tp} \\
S  &= & H P_{k|k-1} H^{\tp} + \frac{1}{|\setZ_{k}|}\\
K  &= & P_{k|k-1} H^{\tp} S^{-1}\\
N^{} &= & S^{-1} \varepsilon \varepsilon^{\tp}
\end{array}
\end{align*}
\rule[0pt]{\columnwidth}{1pt}
\end{table}
\begin{table}[t]
	\caption{Conditional model: smoothing}
	\label{tab:condSmoothing}
	\vspace{-5mm}
	\rule[0pt]{\columnwidth}{1pt}
	\begin{align*}
	\begin{array}{rcl}
	m_{k|K} & = & m_{k|k} + \left(G \otimes \Id \right) \left( m_{k+1|K} - m_{k+1|k}\right) \\
	P_{k|K} & = &P_{k|k} - G\left( P_{k+1|k} - P_{k+1|K} \right) G^{\tp} \\
	v_{k|K} & = & v_{k|k} + \eta^{-1} \left(v_{k+1|K}-v_{k+1|k} - \frac{2(d+1)^{2}}{n}\right) \\
	V_{k|K} & = & V_{k|k} + \eta^{-1} A^{-1} \left(V_{k+1|K} - V_{k|k}\right) (A^{-1})^{\tp} \\
	G & = & P_{k|k} F_k^{\tp} P_{k+1 | k}^{-1} \\
	\eta & = & 1 + \frac{v_{k+1|K}-v_{k|k}-3(d+1)}{n}
	\end{array}
	\end{align*}
	\rule[0pt]{\columnwidth}{1pt}
\end{table}

\subsection{Prediction, update, and smoothing}

The prediction and the update for the conditional model are reproduced in Table~\ref{tab:condPrediction} and in Table~\ref{tab:condUpdate}, respectively. 
The smoothing is given in the following theorem.
\begin{theorem}\label{thm:condSmoothing}
	Let the densities $p(\xi_{k} | \setZ_{1:k})$, $p(\xi_{k+1} | \setZ_{1:K})$ and $p(\xi_{k+1} | \setZ_{1:k})$ be conditional Gaussian inverse Wishart \eqref{eq:conditionalGIW}, and let the transition density be Gaussian Wishart \eqref{eq:condTransitionDensity}. The smoothed density $p(\xi_{k} | \setZ_{1:K})$, see \eqref{eq:SmoothingBackwards}, is conditional Gaussian inverse Wishart, see with parameters $(m_{k|K},P_{k|K},v_{k|K},V_{k|K})$ given in Table~\ref{tab:condSmoothing}.\hfill$\square$
\end{theorem}

The proof of Theorem~\ref{thm:condSmoothing} is given in Appendix~\ref{app:cond_model_smoothing}.


\section{Factorized model}
\label{sec:factSmoothing}

For the random matrix model in \cite{FeldmannFK:2011}, we have factorised Gaussian inverse Wishart densities, cf. \eqref{eq:factorizedGIW},
and under assumed density filtering we seek a smoothed density of the same form, i.e.,
\begin{subequations}
\begin{align}
	p(\xi_{k} | \setZ_{1:K}) = & p\left(\sx_{k}|\setZ_{1:K}\right)p\left(\ext_{k}|\setZ_{1:K}\right) \\
= & \Npdfbig{\sx_{k}}{m_{k|K}}{P_{k|K}} \nonumber \\
& \times \IWishpdf{\ext_{k}}{v_{k|K}}{V_{k|K}}.
\end{align}
\label{eq:factSmoothedGIW}
\end{subequations}

\subsection{Assumptions, approximations}
The following assumption is made for the factorized \giw model, see \cite{FeldmannFK:2011}.
\begin{assumption}\label{ass:IndependentKinematicTransition}
	The time evolution of the kinematic state is independent of the extent state,
	\begin{align}
		p\left(\sx_{k+1}|\ext_{k+1},\sx_{k}\right) = p\left(\sx_{k+1}|\sx_{k}\right)
	\end{align}
	\hfill$\square$
\end{assumption}
The validity of this assumption is discussed in \cite{FeldmannFK:2011,GranstromO:2014}. The transition density is Gaussian Wishart \cite{GranstromO:2014},
\begin{align}
	p(\xi_{k+1}|\xi_{k}) 
= & \Npdfbig{\sx_{k+1}}{f_k \left( \sx_{k} \right)}{Q_k}  \label{eq:factTransitionDensity}\\
& \times \Wishpdf{\ext_{k+1}}{n_{k}}{\frac{M(\sx_{k})\ext_{k} M^{\tp}(\sx_{k})}{n_{k}}} \nonumber
\end{align}
where the function $f_{k}(\cdot) : \mathbb{R}^{n_x} \rightarrow \mathbb{R}^{n_x}$ is the motion model, the $n_x \times n_x$ matrix $Q_k$ is the process noise covariance, the degrees of freedom $n_k \geq d$ govern the uncertainty of the time evolution of the extent, and the function $M(\cdot) : \mathbb{R}^{n_x} \rightarrow \mathbb{I}^{d \times d}$ describes how the extent changes over time due to the object motion. For example, $M(\cdot)$ can be a rotation matrix.  In what follows, we write $M_{\sx} = M(\sx)$ for brevity.

The measurement model is
\begin{align}
	p(\sz | \sx_k,\ext_k) = \Npdfbig{\sz}{\tilde{H}_k\sx_{k}}{\rho\ext_{k}+R_{k}},
\end{align} 
where the $d \times n_x$ matrix $\tilde{H}_k$ is the measurement model, $\rho>0$ is a scaling factor, and $R_{k}\in\mathbb{S}_{+}^{d}$ is the measurement noise covariance. The scaling factor $\rho$ and the noise covariance $R_{k}$ were added to better model scenarios where the sensor noise is large in relation to the size of the extended object, see discussion in \cite[Sec. 3]{FeldmannFK:2011}. In this paper, to enable a straightforward comparison to the conditional model, which assumes that the sensor noise is small in comparison to the size of the extended object, we focus on the case $\rho=1$ and $R_{k} = \mathbf{0}_{d\times d}$.

\subsection{Prediction, update, and smoothing}

The prediction and the update for the conditional model are reproduced in Table~\ref{tab:factPrediction} and in Table~\ref{tab:factUpdate}, respectively. The smoothing is given in the following theorem.
\begin{theorem}\label{thm:factSmoothing}
	Let the densities $p(\xi_{k} | \setZ_{1:k})$, $p(\xi_{k+1} | \setZ_{1:K})$ and $p(\xi_{k+1} | \setZ_{1:k})$ be factorised Gaussian inverse Wishart \eqref{eq:factorizedGIW}, and let the transition density be Gaussian Wishart \eqref{eq:factTransitionDensity}. The smoothed density $p(\xi_{k} | \setZ_{1:K})$, see \eqref{eq:SmoothingBackwards}, is factorised Gaussian inverse Wishart, see with parameters $(m_{k|K},P_{k|K},v_{k|K},V_{k|K})$ given in Table~\ref{tab:factSmoothing}.\hfill$\square$
\end{theorem}

The proof of Theorem~\ref{thm:factSmoothing} is given in Appendix~\ref{app:FactSmoothing}.

\begin{table}
\caption{Factorized model: prediction}
\label{tab:factPrediction}
\vspace{-5mm}
\rule[0pt]{\columnwidth}{1pt}
%
If $M_{\sx} = A$, where $A$ is a $d\times d$ invertible matrix,
\begin{align*}
\begin{array}{rcl}
m_{k+1|k} &= & f_{k}( m_{k|k} ) \\
P_{k+1|k} &= & \tilde{F}_{k}P_{k|k}\tilde{F}_{k}^{\tp} + Q \\
v_{k+1|k} & = & d+1 + \left(1 + \frac{v_{k|k}-2d-2}{n}\right)^{-1}(v_{k|k}-d-1) \\
V_{k+1|k} & = & \left(1 + \frac{v_{k|k}-d-1}{n-d-1}\right)^{-1} A V_{k|k} A^{\tp} \\
\tilde{F}_{k} & = & \left. \nabla_{\sx} f_{k}(\sx) \right |_{\sx=m_{k|k}}
\end{array}
\end{align*}
else,
\begin{align*}
\begin{array}{rcl}
m_{k+1|k} &= & f_{k}( m_{k|k} ) \\
P_{k+1|k} &= & \tilde{F}_{k}P_{k|k}\tilde{F}_{k}^{\tp} + Q \\
v_{k+1|k} & = & d+1 + \eta^{-1}(v_{k|k}-d-1) \\ 
V_{k+1|k} & = & \eta^{-1} \left(1 - \frac{d+1}{s}\right)\left(1 - \frac{d+1}{n}\right) C_{2} \\
\eta & = & 1 + (v_{k|k} -2d-2)\left( \frac{1}{s} + \frac{1}{n} - \frac{d+1}{ns} \right) \\
s & = & \frac{d+1}{d}\tr\left\{ C_{1} C_{2} \left( C_{1} C_{2} - \Id \right)^{-1} \right\} \\
\tilde{F}_{k} & = & \left. \nabla_{\sx} f_{k}(\sx) \right |_{\sx=m_{k|k}} \\
C_{1} & = & \mathbb{E}_{k|k}\left[ \left(M_{\sx} V_{k|k} M_{\sx}^{\tp}\right)^{-1} \right] \\ 
C_{2} & = & \mathbb{E}_{k|k}\left[ M_{\sx} V_{k|k} M_{\sx}^{\tp} \right]
\end{array}
\end{align*}
\rule[0pt]{\columnwidth}{1pt}
\end{table}
\begin{table}
\caption{Factorized model: update}
\label{tab:factUpdate}
\vspace{-5mm}
\rule[0pt]{\columnwidth}{1pt}
\begin{align*}
\begin{array}{rcl}
m_{k|k} &= & m_{k|k-1} + K \varepsilon \\
P_{k|k} &= & P_{k|k-1} -K S K^{\tp}\\
v_{k|k} &= & v_{k|k-1} + |\setZ_{k}|\\
V_{k|k} &= & V_{k|k-1}+\hat{N} + \hat{Z} \\
\varepsilon^{} &= &  \bar{\sz} - \tilde{H} m_{k|k-1}\\
\bar{\sz}  &= &  \frac{1}{|\setZ_{k}|}\sum_{\sz^{i} \in \setZ_{k}}{\sz_{}^{i}} \\
Z_{}^{} &= & \sum_{\sz_{k}^{i}\in \setZ_{k}} \left(\sz_{}^{i} - \bar{\sz}_{}^{}\right)\left(\sz_{}^{i} - \bar{\sz}_{}^{}\right)^{\tp} \\
\hat{X} & = & V_{k|k-1} \left(v_{k|k-1}-2d-2\right)^{-1} \\
Y &= & \rho \hat{X} + R\\
S^{}  &= & \tilde{H} P_{k|k-1} \tilde{H}^{\tp} + \frac{Y}{|\setZ_{k}|}\\
K^{}  &= & P_{k|k-1}\tilde{H}^{\tp} S^{-1}\\
\hat{N} &= & \hat{X}^{\frac{1}{2}} S^{-\frac{1}{2}}  \varepsilon \varepsilon^{\tp} S^{-\frac{\tp}{2}} \hat{X}^{\frac{\tp}{2}} \\
\hat{Z} & = & \hat{X}^{\frac{1}{2}} Y^{-\frac{1}{2}}  Z Y^{-\frac{\tp}{2}} \hat{X}^{\frac{\tp}{2}}
\end{array}
\end{align*}
\rule[0pt]{\columnwidth}{1pt}
\end{table}

\begin{table}
	\caption{Factorized model: smoothing}
	\label{tab:factSmoothing}
	\vspace{-5mm}
	\rule[0pt]{\columnwidth}{1pt}
	If $M_{\sx} = A$, where $A$ is a $d\times d$ invertible matrix,
	\begin{align*}
	\begin{array}{rcl}
	m_{k|K} & = & m_{k|k} + G_{k} \left( m_{k+1|K} - m_{k+1|k}\right) \\
	P_{k|K} & = & P_{k|k} - G \left( P_{k+1|k} - P_{k+1|K} \right) G^{\tp} \\
	v_{k|K} & = & v_{k|k} + \eta^{-1} \left(v_{k+1|K} - v_{k+1|k} -\frac{2(d+1)^2}{n}\right) \\
	V_{k|K} & = & V_{k|k} + \eta^{-1} A^{-1}\left(V_{k+1|K} - V_{k+1|k}\right) (A^{-1})^{\tp} \\ 
	G & = & P_{k|k} \tilde{F}_k^{\tp} P_{k+1 | k}^{-1} \\
	\eta & = & 1 + \frac{v_{k+1|K} - v_{k+1|k}-3(d+1)}{n}
	\end{array}
	\end{align*}
	else
	\begin{align*}
	\begin{array}{rcl}
	m_{k|K} & = & m_{k|k} + G_{k} \left( m_{k+1|K} - m_{k+1|k}\right) \\
	P_{k|K} & = & P_{k|k} - G \left( P_{k+1|k} - P_{k+1|K} \right) G^{\tp} \\
	v_{k|K} & = & v_{k|k} + \eta_{2}^{-1} \left(g -\frac{2(d+1)^2}{h+d+1}\right) \\
	V_{k|K} & = & V_{k|k} + \eta_{3}^{-1}C_{4} \\ 
	G & = & P_{k|k} \tilde{F}_k^{\tp} P_{k+1 | k}^{-1} \\
	W & = & V_{k+1|K} - V_{k+1|k} \\
	w & = & v_{k+1|K} - v_{k+1|k} \\
	g & = & \eta_{1}^{-1}\left(w - \frac{2(d+1)^2}{n}\right) \\
	h & = & \frac{d+1}{d}\tr\left\{ C_{3} C_{4} \left( C_{3} C_{4} - \Id \right)^{-1} \right\} \\
	\eta_{1} & = & 1+\frac{w-3(d+1)}{n} \\
	\eta_{2} & = & 1+\frac{g-3d-3}{h+d+1} \\
	\eta_{3} & = & 1 + \frac{g-d-1}{h-d-1} \\
	C_{3} & = & \mathbb{E}_{k|K}\left[ \left(M_{\sx}^{-1} W  (M_{\sx}^{-1})^{\tp}\right)^{-1} \right] \\
		& = & \mathbb{E}_{k|K}\left[ M_{\sx}^{\tp} W^{-1} M_{\sx} \right] \\
	C_{4} & = & \mathbb{E}_{k|K}\left[ M_{\sx}^{-1} W (M_{\sx}^{-1})^{\tp} \right] \\
		& = & \mathbb{E}_{k|K}\left[ \left( M_{\sx}^{\tp} W^{-1} M_{\sx} \right)^{-1}\right]
	\end{array}
	\end{align*}
	\rule[0pt]{\columnwidth}{1pt}
\end{table}

\subsection{Expected value approximation}

Note that both the prediction and the smoothing require expected values, see $C_{1}$ and $C_{2}$ in Table~\ref{tab:factPrediction}, and $C_{3}$ and $C_{4}$ in Table~\ref{tab:factSmoothing}. For a Gaussian distributed vector $\sx \sim \mathcal{N}(m,P)$, the expected value of $M_{\sx}VM_{\sx}^{\tp}$
 can be approximated using third order Taylor expansion,
\begin{align}
C_{1} \approx & \left( M(m)VM(m)^{\tp} \right)^{-1} \nonumber \\
& + \sum_{i=1}^{n_{x}} \sum_{j=1}^{n_{x}} \left.\frac{\diff^{2} \left( M_{\sx}VM_{\sx}^{\tp} \right)^{-1}}{\diff \sx^{[i]} \diff \sx^{[j]}}\right|_{\sx=m} P^{[i,j]}
\end{align}
where $\sx^{[i]}$ is the $i$th element of $\sx$, $P^{[i,j]}$ is the $i,j$th element of $P$. The necessary differentiations are
\begin{subequations}
\begin{align}
	\frac{\diff M_{\sx}VM_{\sx}^{\tp}}{\diff \sx^{[j]}} = &   \frac{\diff M_{\sx}}{\diff \sx^{[j]}}V M_{\sx}^{\tp} + M_{\sx} V \frac{\diff M_{\sx}^{\tp}}{\diff \sx^{[j]}} \\
	\frac{\diff^{2} M_{\sx}VM_{\sx}^{\tp}}{\diff \sx^{[i]} \diff \sx^{[j]}}  = &  \frac{\diff^{2} M_{\sx}}{\diff \sx^{[i]} \diff \sx^{[j]}}VM_{\sx}^{\tp} + \frac{\diff M_{\sx}}{\diff \sx^{[j]}}V\frac{\diff M_{\sx}^{\tp}}{\diff \sx^{[i]}} \nonumber \\
	&  + \frac{\diff M_{\sx}}{\diff \sx^{[i]}}V\frac{\diff M_{\sx}^{\tp}}{\diff \sx^{[j]}} + M_{\sx} V \frac{\diff^{2} M_{\sx}^{\tp}}{\diff \sx^{[i]} \diff \sx^{[j]}}
\end{align}
and, for any function $N_{\sx}=N(\sx)$,
\begin{align}
	\frac{\diff^{2} N_{\sx}^{-1}}{\diff \sx^{[i]} \diff \sx^{[j]}} = & N_{\sx}^{-1} \frac{\diff N_{\sx}}{\diff \sx^{[j]}} N_{\sx}^{-1}\frac{\diff N_{\sx}}{\diff \sx^{[i]}} N_{\sx}^{-1} - N_{\sx}^{-1} \frac{\diff^{2} N_{\sx}}{\diff\sx^{[i]} \diff\sx^{[j]}} N_{\sx}^{-1} \nonumber \\
	&  +N_{\sx}^{-1} \frac{\diff N_{\sx}}{\diff \sx^{[i]}} N_{\sx}^{-1}\frac{\diff N_{\sx}}{\diff \sx^{[j]}} N_{\sx}^{-1}
\end{align}
\end{subequations}
 The expected values $C_{2}$, $C_{3}$ and $C_{4}$ can be approximated analogously.


\section{Simulation study}
\label{sec:simulationStudy}
In this section we present the results of a simulation study. In all simulations, the dimension of the extent is $d=2$.

\subsection{Implemented smoothers}
Three different smoothers were implemented.
\subsubsection{Conditional \giw model with constant velocity motion model (CCV)} The state vector contains 2D Cartesian position and velocity, $\sx_{k} = [p_{k}^{x}, \ p_{k}^{y},\ v_{k}^{x}, \ v_{k}^{y}]^{\tp}$, $n_x=4$ and $s=2$. The following models are used,
		\begin{subequations}
		\begin{align}
			F_{k} &= \begin{bmatrix} 1 & T \\ 0 & 1\end{bmatrix}, \label{eq:CVmodel} \\
			D_{k} &= \sigma_{a}^{2} \begin{bmatrix} \frac{T^4}{4} & \frac{T^3}{2} \\ \frac{T^3}{2} & T^2 \end{bmatrix}, \label{eq:CVnoiseCovariance} \\
			H_{k} &= \begin{bmatrix} 1 & 0 \end{bmatrix}.
		\end{align}
		\end{subequations}
		$A = \Id$, and $n_k = 100$, where $T$ is the sampling time.
		
\subsubsection{Factorized \giw model with constant velocity motion model (FCV)} The state vector contains 2D Cartesian position and velocity, $\sx_{k} = [p_{k}^{x}, \ p_{k}^{y},\ v_{k}^{x}, \ v_{k}^{y}]^{\tp}$, and $n_x=4$. The following models are used,
		\begin{subequations}
		\begin{align}
			f_{k}(\sx) &= \begin{bmatrix}  \mathbf{I}_{2} & T \mathbf{I}_{2} \\ \mathbf{0}_{2\times 2} &  \mathbf{I}_{2} \end{bmatrix} \sx, \\ 
			Q_{k} &=  \sigma_{a}^{2} \begin{bmatrix} \frac{T^4}{4} \mathbf{I}_{2} & \frac{T^3}{2} \mathbf{I}_{2} \\ \frac{T^3}{2} \mathbf{I}_{2} & T^2 \mathbf{I}_{2} \end{bmatrix} \\
			\tilde{H}_{k} & = \begin{bmatrix}  \mathbf{I}_{2} & \mathbf{0}_{2\times 2} \end{bmatrix} 
		\end{align}
		\end{subequations}
		$n_k = 100$, and $M_{\sx} = \Id$.
		
\subsubsection{Factorized \giw model with coordinated turn motion model (FCT)} The state vector contains 2D Cartesian position and velocity, as well as turn-rate, $\sx_{k} = [p_{k}^{x}, \ p_{k}^{y},\ v_{k}^{x}, \ v_{k}^{y}, \ \omega_{k}]^{\tp}$, and $n_x=5$. The following models are used,
		\begin{subequations}
		\begin{align}
			f_{k}(\sx_k) & = {\scriptsize \begin{bmatrix} 
				1 & 0 & \frac{\sin(T\omega_k)}{\omega_k} & -\frac{1-\cos(T\omega_k)}{\omega_k} & 0 \\
				0 & 1 & \frac{1-\cos(T\omega_k)}{\omega_k} & \frac{\sin(T\omega_k)}{\omega_k} & 0 \\
				0 & 0 & \cos(T\omega_k) & - \sin(T\omega_k) & 0 \\
				0 & 0 &  \sin(T\omega_k) & \cos(T\omega_k) & 0 \\
				0 & 0 & 0 & 0 & 1
				\end{bmatrix} } \sx_k, \label{eq:CTmodel} \\
			Q_{k} &= G \diag{[ \sigma_{a}^{2},\  \sigma_{a}^{2}, \ \sigma_{\omega}^{2} ]} G^{\tp}, \label{eq:CTnoiseCovariance} \\
			G  & = \begin{bmatrix} \frac{T^2}{2} \mathbf{I}_{2} & \mathbf{0}_{2\times 1} \\ T\mathbf{I}_{2} & \mathbf{0}_{2\times 1} \\ \mathbf{0}_{1\times 2} & 1 \end{bmatrix} \\
			M_{\sx} & = \begin{bmatrix} \cos(T\omega) & -\sin(T\omega) \\ \sin(T\omega) & \cos(T\omega) \end{bmatrix},\\
			\tilde{H}_{k} & = \begin{bmatrix}  \mathbf{I}_{2} & \mathbf{0}_{2\times 3} \end{bmatrix} 
		\end{align}
		\end{subequations}
		and $n_k = \infty$. For the matrix transformation function $M_{\sx}$ we have the following,
		\begin{subequations}
		\begin{align}
			& M_{\sx}^{-1} = M_{\sx}^{\tp} \\ 
			& \frac{\diff M_{\sx}}{\diff \sx^{[i]}} = \begin{cases} \scriptsize  T \begin{bmatrix} -\sin(T\omega) & -\cos(T\omega) \\ \cos(T\omega) & -\sin(T\omega) \end{bmatrix} & i=5 \\ \mathbf{0}_{2\times 2} & i \neq 5 \end{cases} \\
			& \frac{\diff^2 M_{\sx}}{\diff \sx^{[i]} \diff \sx^{[j]}} = \begin{cases} \scriptsize  T^2 \begin{bmatrix} -\cos(T\omega) & \sin(T\omega) \\ -\sin(T\omega) & -\cos(T\omega) \end{bmatrix} & i,j=5 \\ \mathbf{0}_{2\times 2} & i,j \neq 5 \end{cases}
		\end{align}
		\end{subequations}
		
\subsection{Simulated scenarios}

We focused on two types of scenarios: in the first the true tracks were generated by a constant velocity model; in the second the true tracks were generated by a coordinated turn model. This allows us to test the different smoothers both when their respective motion models match the true model, and when there is motion model mis-match.

The CV tracks were generated using the CV model \eqref{eq:CVmodel} and \eqref{eq:CVnoiseCovariance} with $\sigma_{a} = 1$; the extent's major axis was simulated to be aligned with the velocity vector of the extended object. 
The CT tracks were generated using the CT model \eqref{eq:CTmodel} and \eqref{eq:CTnoiseCovariance} with $\sigma_{a} = 1$ and $\sigma_{\omega} = \pi/180$; the extent's major axis was simulated to be aligned with the velocity vector of the extended object. 

For both motion models, in each time step $k$, a detection process was simulated by first sampling a probability of detection $p_{\rm D}$, and, if the object is detected, sampling $N_{z}$ detections using a Gaussian likelihood. We simulated two combinations: $(p_{\rm D}, \ N_{z}) = (0.25, \ 10)$ and $(p_{\rm D}, \ N_{z}) = (0.75, \ 10)$.

\subsection{Performance evaluation}

For performance evaluation of extended object estimates with ellipsoidal extents, a comparison study has shown that among six compared performance measures, the Gaussian Wassterstein Distance (\gwd) metric is the best choice \cite{YangBG:2016}. The \gwd is defined as \cite{GivensS:1984}
\begin{align}
	\Delta_{k|\ell} = & \| \mathbf{p}_{k} - \hat{\mathbf{p}}_{k|\ell} \|^{2} \\
	& + \tr\left(\ext_{k}+\hat{\ext}_{k|\ell}-2 \left( \ext_{k}^{\frac{1}{2}} \hat{\ext}_{k|\ell} \ext_{k}^{\frac{1}{2}} \right)^{\frac{1}{2}} \right), \nonumber
\end{align}
where the expected extended object state is
\begin{align}
	\hat{\xi}_{k|\ell} = & \mathbb{E}_{p(\xi_{k} | \setZ_{1:\ell})} \left[ \xi_{k} \right] = \left(\hat{\sx}_{k|\ell},\ \hat{\ext}_{k|\ell}\right) \\
	= & \left( m_{k|\ell} , \ \frac{V_{k|\ell}}{v_{k|\ell} - 2d-2} \right)
\end{align}
and $\mathbf{p}_{k}$ is the position part of the extended object state vector $\sx_{k}$. 

\subsection{Results}

\begin{figure*}[htbp]
	\begin{center}
		\includegraphics[width=\columnwidth]{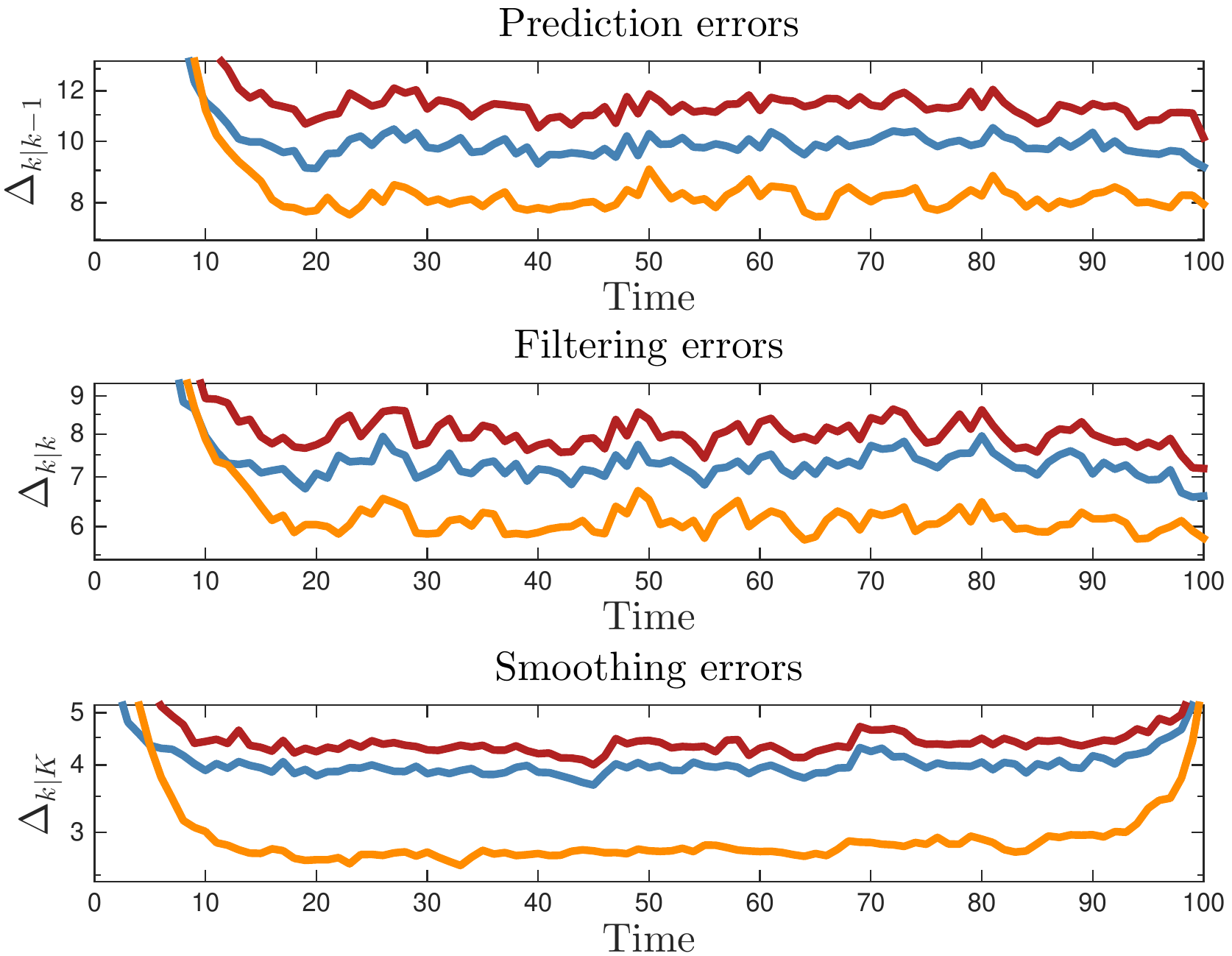}
		\includegraphics[width=\columnwidth]{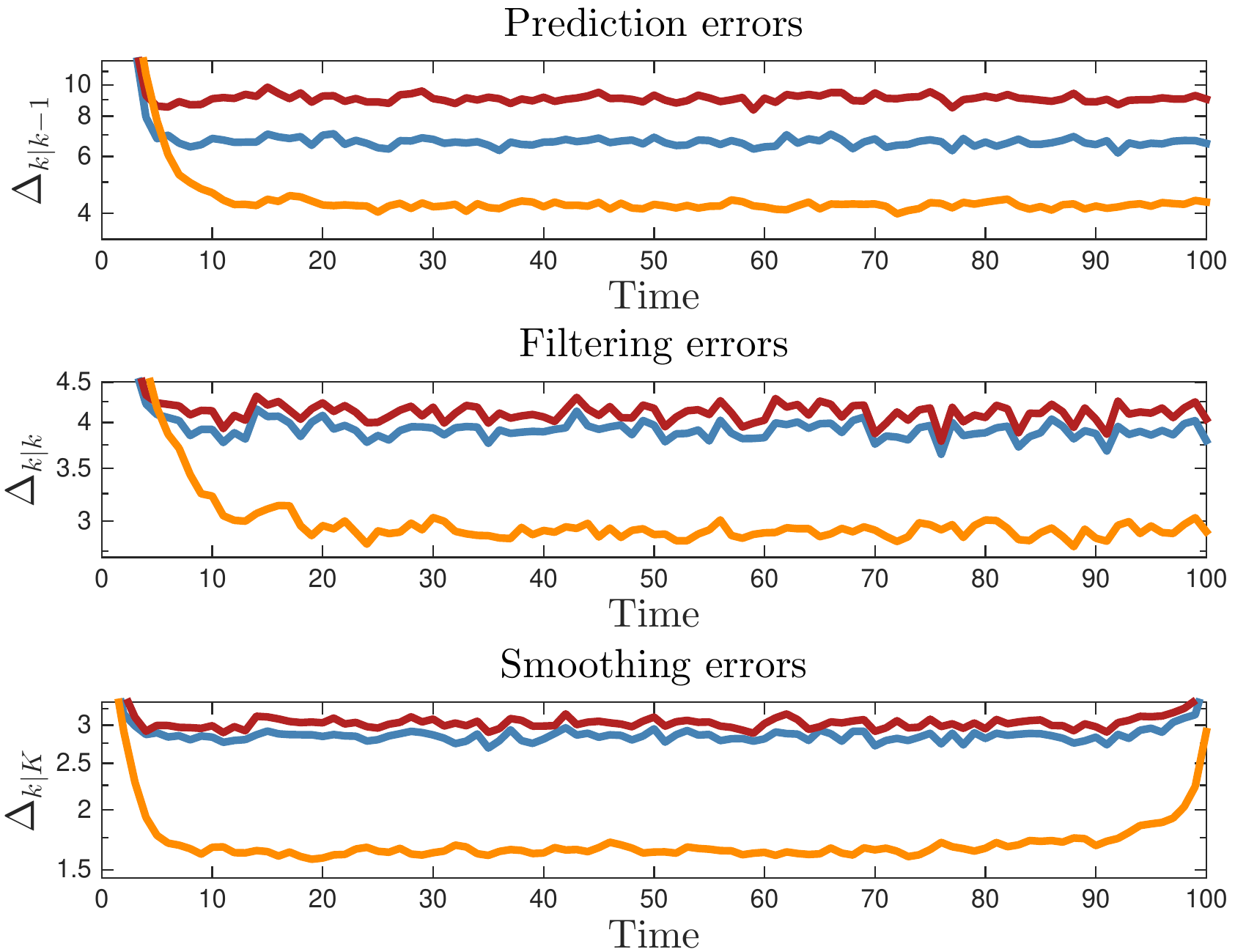}
		\caption{Results for object tracks generated by a CV model, with probability of detection $p_{\rm D}=0.25$ (left) and $p_{\rm D}=0.75$ (right) and, if detected, $N_{z}=10$ object measurements (both). CCV is shown in blue, FCV is shown in red, and FCT is shown in orange. Each line is the median from $1000$ Monte Carlo simulations.}
		\label{fig:GWD_CV}
	\end{center}
\end{figure*}

\begin{figure*}[htbp]
	\begin{center}
		\includegraphics[width=\columnwidth]{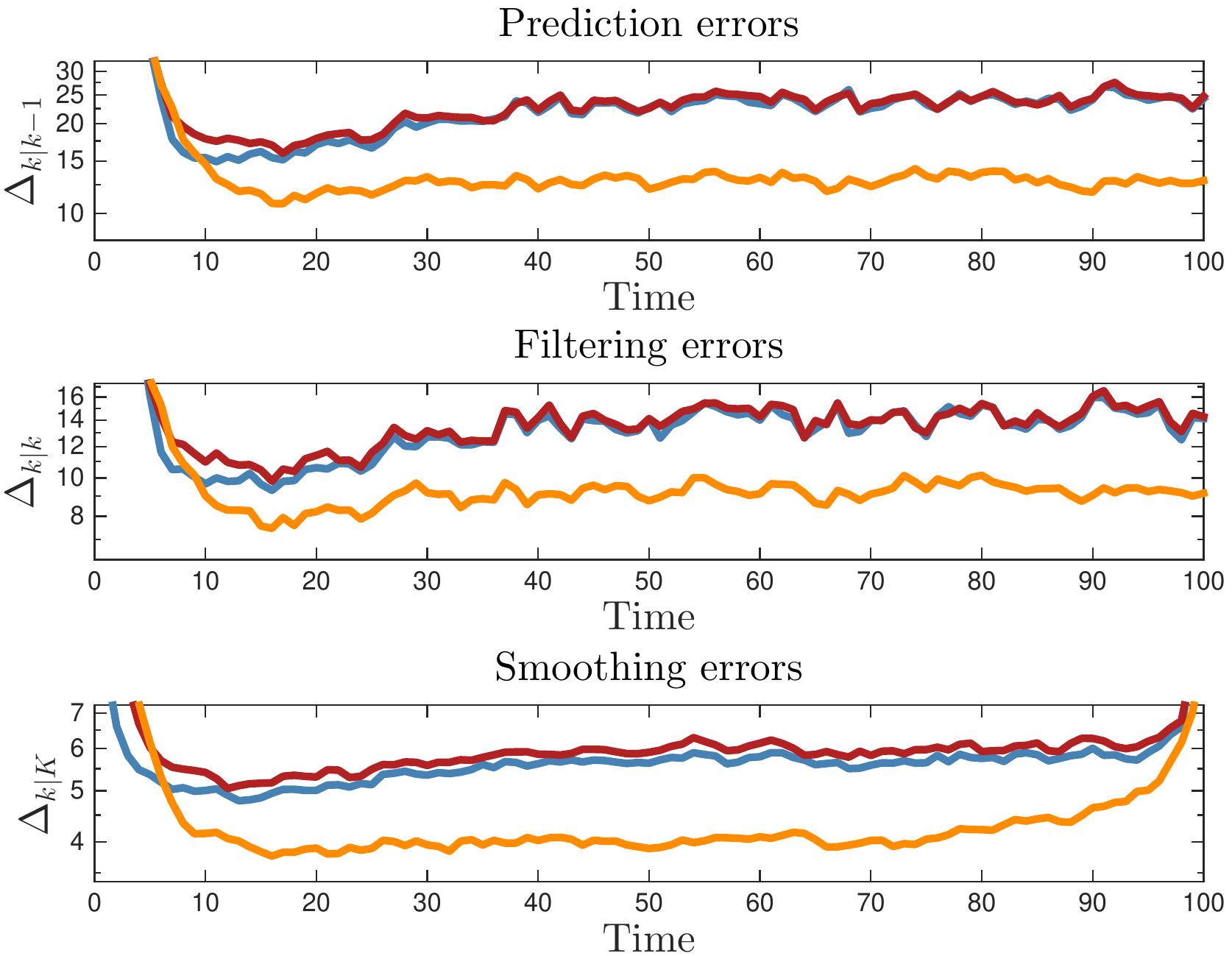}
		\includegraphics[width=\columnwidth]{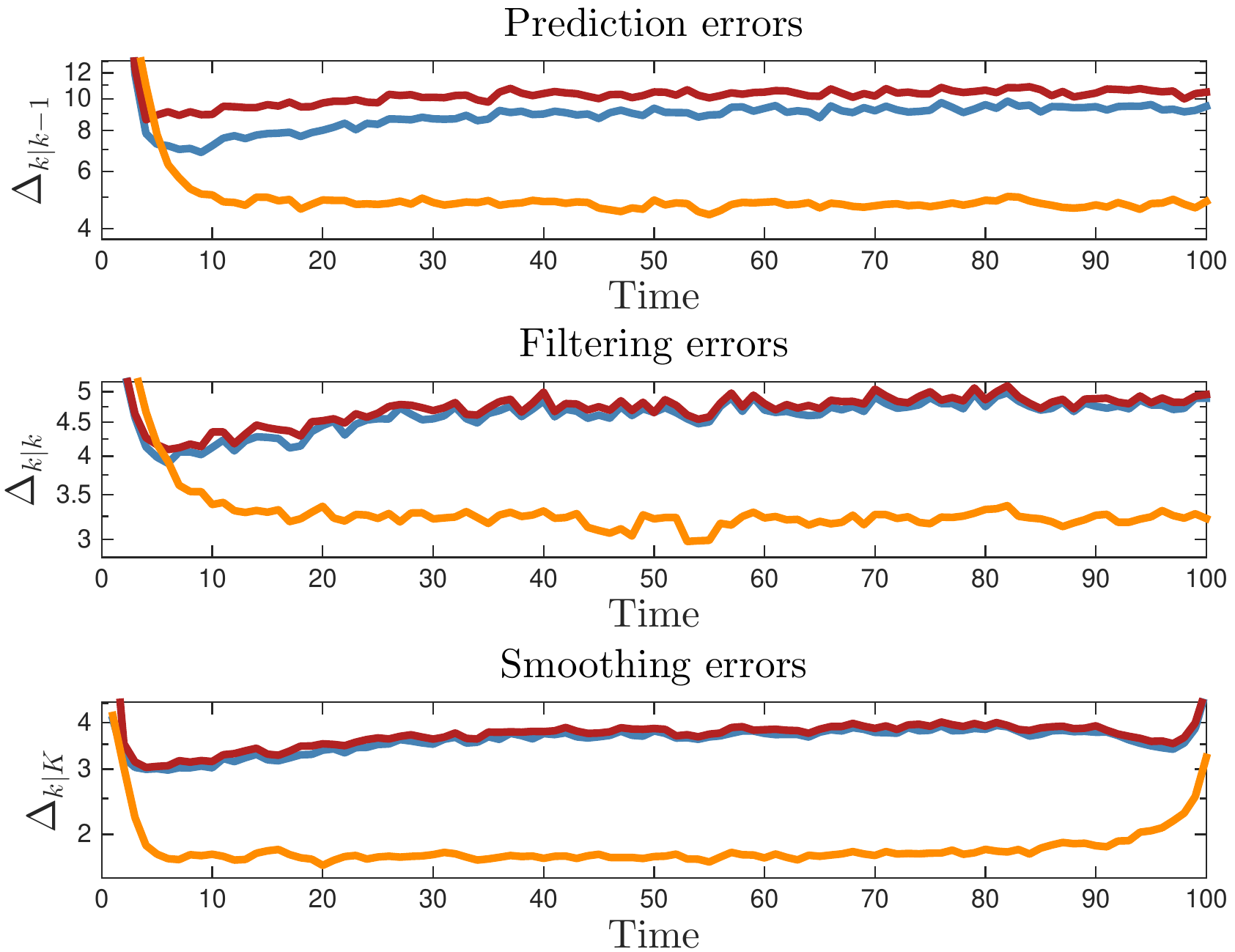}
		\caption{Results for object tracks generated by a CT model, with probability of detection $p_{\rm D}=0.25$ (left) and $p_{\rm D}=0.75$ (right) and, if detected, $N_{z}=10$ object measurements (both). CCV is shown in blue, FCV is shown in red, and FCT is shown in orange. Each line is the median from $1000$ Monte Carlo simulations.}
		\label{fig:GWD_CT}
	\end{center}
\end{figure*}

We show results for estimates $\hat{\xi}_{k|\ell}$ for $\ell \in \{k-1, \ k, \ K\}$, i.e., prediction, filtering and smoothing. Results for true tracks generated by a CV model are shown in Figure~\ref{fig:GWD_CV}; results for true tracks generated by a CT model are shown in Figure~\ref{fig:GWD_CT}.

We see that in all cases, as expected, the smoothing errors are smaller than the filtering errors, which are smaller than the prediction errors. This confirms that the derived smoothers work as they should. It is also in accordance with expectations that the performance is worse when the probability of detection is lower. Perhaps counter-intuitive is that for CV true motion, the FCT smoother performs best despite the modelling error in the motion model. We believe that this is due, at least in part, to the standard assumption that the orientation of the extent ellipse is aligned with the velocity vector. The motion noice on the velocity vector introduces rotations on the extent ellipse, and the motion model used in FCT captures these rotations better.

\section{Conclusions and future work}
\label{sec:Conclusions}

This paper presented Bayesian smoothing for the random matrix model used in extended object tracking. Two variants of Gaussian inverse Wishart state densities exist in the literature, a conditional and a factorised; closed form Bayesian smoothing was derived for both of them. The derived smoothers were implemented and tested in a simulation scenario. In future work the smoothers will be used with real data, e.g., data from camera, lidar or radar.




\appendix

\subsection{Preliminary results}

In this appendix we present some preliminary results that are used in the proofs of Theorems~\ref{thm:condSmoothing} and \ref{thm:factSmoothing}. The first two Lemmas regard products, and ratios, of inverse Wishart pdfs, respectively.
\begin{lemma}\label{lem:ProductInverseWishart}
The product of two inverse Wishart pdfs is proportional to an inverse Wishart pdf,
\begin{align}
	& \IWishpdf{\ext}{a}{A}\IWishpdf{X}{b}{B} \nonumber \\
	& \propto \IWishpdf{X}{a+b}{A+B}
\end{align}
\hfill$\square$
\end{lemma}
\emph{Proof:} This follows from the definition of the inverse Wishart pdf.
\begin{lemma}\label{lem:FractionInverseWishart}
The fraction of two inverse Wishart pdfs is proportional to an inverse Wishart pdf,
\begin{align}
	& \frac{\IWishpdf{\ext}{a}{A}}{\IWishpdf{X}{b}{B}}  \propto \IWishpdf{X}{a-b}{A-B}
\end{align}
\hfill$\square$
\end{lemma}
\emph{Proof:} This follows from the definition of the inverse Wishart pdf.

The following two Lemmas are related to the use Wishart transition densities and inverse Wishart state densities for the extent matrix.
\begin{lemma}\label{lem:WishartInverseWishart}
For Wishart and inverse Wishart pdfs, the following holds,
\begin{align}
	& \Wishpdf{Y}{n}{\frac{M\ext M^{\tp}}{n}} \nonumber \\
	& \propto \IWishpdf{\ext}{n}{nM^{-1}Y\left(M^{-1}\right)^{\tp}}.
\end{align}
\hfill$\square$
\end{lemma}
\emph{Proof:} This follows from the definitions of the Wishart pdf and the inverse Wishart pdf.
\begin{lemma}\label{lem:Integral_condW_IW}
	\begin{align}
	& \int \Wishpdf{\ext}{v}{V} \IWishpdf{V}{w}{W} \diff V \nonumber \\
	= & \mathcal{GB}_{d}^{II}\left(\ext ; \frac{v}{2}, \frac{w-d-1}{2}, W, \mathbf{0}_{d \times d}\right)
	\end{align}
\end{lemma}
\emph{Proof:} see \cite[Prob. 5.33]{GuptaN:2000}.
\begin{lemma}\label{lem:Integral_condIW_W}
	\begin{align}
	& \int \IWishpdf{\ext}{v}{V} \Wishpdf{V}{w}{W} \diff V \nonumber \\
	= & \mathcal{GB}_{d}^{II}\left(\ext ; \frac{w}{2}, \frac{v-d-1}{2}, W, \mathbf{0}_{d\times d}\right)
	\end{align}
\end{lemma}
\emph{Proof:} see \cite[Thm. 3]{GranstromO:2014}.


For approximations of densities, the Kullback-Leibler divergence (\kldiv) is often minimised to find the optimal approximation. The following Lemma is about approximation by a factorised density.
\begin{lemma}\label{lem:KLminFactorisedApproximation}
	For two random variables $\sx$ and $\ext$, with joint density $p(\sx,\ext)$, the factorised density $q^{\star}(\sx)q^{\star}(\ext)$ that minimises the \kldiv to $p(\sx,\ext)$,
	\begin{align}
		q^{\star}(\sx)q^{\star}(\ext) = \arg\min_{q(\sx)q(\ext)} {\rm KL}\left(p(\sx,\ext) || q(\sx)q(\ext) \right)
	\end{align}
	is given by the marginals,
	\begin{align}
		q^{\star}(\sx) & = \int p(\sx,\ext) \diff \ext \\
		q^{\star}(\ext) & = \int p(\sx,\ext) \diff \sx
	\end{align}
	\hfill$\square$
\end{lemma}
\emph{Proof:} This is a previously know result that follows from the definition of the Kullback-Leibler divergence.

Approximation of matrix valued densities as Wishart, or inverse Wishart, densities, by minimisation of the \kldiv, is presented in \cite[Thm. 1]{GranstromO:2014} and \cite[Thms. 3 \& 4]{GranstromO:2013spawn}. The \kldiv minimisation leads to matching of the expected logarithm of the determinant of the extent matrix, as well as either the expected extent matrix, or the expected inverse extent matrix. A simpler closed form approximation is obtained if instead the expected random matrix and the expected inverse are match, i.e., not matching the expected log-determinant; this is shown in the following four Lemmas.
\begin{lemma}\label{lem:Approx_IW_as_W}
	By matching the expected values $E[\ext]$ and $E[\ext^{-1}]$, the inverse Wishart density $\IWishpdf{\ext}{v}{V}$ can be approximated by a Wishart density $\Wishpdf{\ext}{w}{W}$ with parameters
	\begin{align}
		w & = v-d-1 \\
		W & =  \frac{V}{(v-2d-2)(v-d-1)}
	\end{align}
	\hfill$\square$
\end{lemma}
\emph{Proof:} this follows from the definitions of the expected values, see \cite{GuptaN:2000}.
\begin{lemma}\label{lem:Approx_W_as_IW}
	By matching the expected values $E[\ext]$ and $E[\ext^{-1}]$, the Wishart density $\IWishpdf{\ext}{w}{W}$ can be approximated by an inverse Wishart density $\Wishpdf{\ext}{v}{V}$ with parameters
	\begin{align}
		v & = w+d+1 \\
		V & =  Ww(w-d-1)
	\end{align}
	\hfill$\square$
\end{lemma}
\emph{Proof:} this follows from the definitions of the expected values, see \cite{GuptaN:2000}.
\begin{lemma}\label{lem:Approx_GB2_as_W}
	By matching the expected values $E[\ext]$ and $E[\ext^{-1}]$, the generalized Beta type 2 density $\mathcal{GB}_{d}^{II}\left(\ext ; \frac{a}{2}, \frac{b}{2}, A, \mathbf{0}_{d\times d}\right)$ can be approximated by a Wishart density $\Wishpdf{\ext}{w}{W}$ with parameters
	\begin{align}
		w & = \frac{ab}{a+b-d-1} \\
		W & =  \frac{(a+b-d-1)A}{b(b-d-1)}
	\end{align}
	\hfill$\square$
\end{lemma}
\emph{Proof:} this follows from the definitions of the expected values, see \cite{GuptaN:2000}.
\begin{lemma}\label{lem:Approx_GB2_as_IW}
	By matching the expected values $E[\ext]$ and $E[\ext^{-1}]$, the generalized Beta type 2 density $\mathcal{GB}_{d}^{II}\left(\ext ; \frac{a}{2}, \frac{b}{2}, A, \mathbf{0}_{d\times d}\right)$ can be approximated by an inverse Wishart density $\Wishpdf{\ext}{v}{V}$ with parameters
	\begin{align}
		v & = \frac{ab}{a+b-d-1}+d+1 \nonumber \\
		& = \frac{(a+d+1)(b+d+1)-2(d+1)^2}{a+b-d-1} \\
		V & = \frac{a(a-d-1)}{a+b-d-1}A
	\end{align}
	\hfill$\square$
\end{lemma}
\emph{Proof:} this follows from the definitions of the expected values, see \cite{GuptaN:2000}.


\subsection{Conditional model smoothing}
\label{app:cond_model_smoothing}

For conditional densities \eqref{eq:conditionalDensityGeneral} and the transition density \eqref{eq:condTransitionDensityGeneral}, under Assumptions~\ref{ass:ExtentTransitionDensity} and \ref{ass:SlowlyChangingExtent}, the Bayesian smoothing \eqref{eq:SmoothingBackwards} leads to a conditional smoothed density 
	\begin{subequations}
		\begin{align}
			& p(\xi_{k} | \setZ_{1:K}) = p\left(\sx_{k}|\ext_{k},\setZ_{1:K}\right) p\left(\ext_{k}|\setZ_{1:K}\right)
		\end{align}
where
		\begin{align}
			& p\left(\sx_{k}|\ext_{k},\setZ_{1:K}\right) = p\left(\sx_{k}|\ext_{k},\setZ_{1:k}\right) \nonumber \\
			& \quad \times \int \frac{ p\left(\sx_{k+1}|\sx_{k},\ext_{k}\right)  p\left(\sx_{k+1}|\ext_{k},\setZ_{1:K}\right) }{ p\left(\sx_{k+1}|\ext_{k},\setZ_{1:k}\right) } \diff \sx_{k+1} \\
			& p\left(\ext_{k}|\setZ_{1:K}\right) = p\left(\ext_{k}|\setZ_{1:k}\right) \nonumber \\
			& \quad \times \int \frac{ p\left(\ext_{k+1}|\ext_{k}\right) p\left(\ext_{k+1}|\setZ_{1:K}\right) }{ p\left(\ext_{k+1}|\setZ_{1:k}\right) } \diff \ext_{k+1}
		\end{align}%
		\label{eq:ConditionalModelSeparateSmoothing}%
	\end{subequations}%
The proof of \eqref{eq:ConditionalModelSeparateSmoothing} is given in \eqref{eq:ConditionalModelSeparateSmoothingProof}. 
\begin{figure*}[t]
	\rule{\textwidth}{1pt}
	\begin{subequations}
		\begin{align}
			p&(\xi_{k} | \setZ_{1:K}) = p(\xi_{k} | \setZ_{1:k}) \int \frac{p(\xi_{k+1}|\xi_{k}) p(\xi_{k+1}|\setZ_{1:K})}{p(\xi_{k+1}|\setZ_{1:k})} \diff \xi_{k+1}  \\
			& \overset{\text{A\ref{ass:ExtentTransitionDensity}}}{\approx} p(\xi_{k} | \setZ_{1:k}) \int \frac{ p\left(\sx_{k+1}|\sx_{k},\ext_{k+1}\right) p\left(\ext_{k+1}|\ext_{k}\right) p\left(\sx_{k+1}|\ext_{k+1},\setZ_{1:K}\right)p\left(\ext_{k+1}|\setZ_{1:K}\right) }{ p\left(\sx_{k+1}|\ext_{k+1},\setZ_{1:k}\right)p\left(\ext_{k+1}|\setZ_{1:k}\right) } \diff \xi_{k+1} \\
			& \overset{\text{A\ref{ass:SlowlyChangingExtent}}}{\approx} p(\xi_{k} | \setZ_{1:k}) \int \frac{ p\left(\sx_{k+1}|\sx_{k},\ext_{k}\right) p\left(\ext_{k+1}|\ext_{k}\right) p\left(\sx_{k+1}|\ext_{k},\setZ_{1:K}\right)p\left(\ext_{k+1}|\setZ_{1:K}\right) }{ p\left(\sx_{k+1}|\ext_{k},\setZ_{1:k}\right)p\left(\ext_{k+1}|\setZ_{1:k}\right) } \diff \xi_{k+1} \\
			& = p\left(\sx_{k}|\ext_{k},\setZ_{1:k}\right)p\left(\ext_{k}|\setZ_{1:k}\right) \int \frac{ p\left(\sx_{k+1}|\sx_{k},\ext_{k}\right)  p\left(\sx_{k+1}|\ext_{k},\setZ_{1:K}\right) }{ p\left(\sx_{k+1}|\ext_{k},\setZ_{1:k}\right) } \diff \sx_{k+1} \int \frac{ p\left(\ext_{k+1}|\ext_{k}\right) p\left(\ext_{k+1}|\setZ_{1:K}\right) }{ p\left(\ext_{k+1}|\setZ_{1:k}\right) } \diff \ext_{k+1} \\
			& =  p\left(\sx_{k}|\ext_{k},\setZ_{1:k}\right) \int \frac{ p\left(\sx_{k+1}|\sx_{k},\ext_{k}\right)  p\left(\sx_{k+1}|\ext_{k},\setZ_{1:K}\right) }{ p\left(\sx_{k+1}|\ext_{k},\setZ_{1:k}\right) } \diff \sx_{k+1} p\left(\ext_{k}|\setZ_{1:k}\right) \int \frac{ p\left(\ext_{k+1}|\ext_{k}\right) p\left(\ext_{k+1}|\setZ_{1:K}\right) }{ p\left(\ext_{k+1}|\setZ_{1:k}\right) } \diff \ext_{k+1} \\
			& = p\left(\sx_{k}|\ext_{k},\setZ_{1:K}\right) p\left(\ext_{k}|\setZ_{1:K}\right)
		\end{align}
		\label{eq:ConditionalModelSeparateSmoothingProof}
	\end{subequations}
	\rule{\textwidth}{1pt}
\end{figure*}
We get the following smoothed conditional \giw density
\begin{subequations}
\begin{align}
	p(\sx_{k}|\ext_{k},\setZ_{1:K}) = & \Npdfbig{\sx_{k}}{m_{k|K}}{P_{k|K}\otimes\ext_{k}} \label{eq:smoothed_conditional_kinematic_density} \\
	p(\ext_{k} | \setZ_{1:K})= &\IWishpdf{\ext_{k}}{v_{k|K}}{V_{k|K}} \label{eq:smoothed_conditional_extent_density}
\end{align}
\end{subequations}
with the parameters given in Table~\ref{tab:condSmoothing}. The proof of \eqref{eq:smoothed_conditional_kinematic_density} is simple; the details follow the proof of the RTS-smoother, see, e.g., \cite[Thm. 8.2]{Sarkka:2013}. The proof of \eqref{eq:smoothed_conditional_extent_density} is given in \eqref{eq:ExtentSmoothing}.

\begin{figure*}[t]
\rule{\textwidth}{1pt}
\begin{subequations}
\begin{align}
	p&(\ext_{k} | \setZ_{1:K})  = \IWishpdf{\ext_{k}}{v_{k|k}}{V_{k|k}}  \int \frac{ \Wishpdf{\ext_{k+1}}{n_{k}}{\frac{\ext_{k}}{n_{k}}} \IWishpdf{\ext_{k+1}}{v_{k+1|K}}{V_{k+1|K}}}{ \IWishpdf{\ext_{k+1}}{v_{k+1|k}}{V_{k+1|k}} } \diff\ext_{k+1} \\
	& \overset{\text{L\ref{lem:FractionInverseWishart}}}{\propto} \IWishpdf{\ext_{k}}{v_{k|k}}{V_{k|k}}  \int \Wishpdf{\ext_{k+1}}{n_{k}}{\frac{\ext_{k}}{n_{k}}} \IWishpdf{\ext_{k+1}}{\tilde{v}_{k+1}}{\tilde{V}_{k+1}} \diff\ext_{k+1} \\
	& \overset{\text{L\ref{lem:WishartInverseWishart}}}{=} \IWishpdf{\ext_{k}}{v_{k|k}}{V_{k|k}}  \int \IWishpdf{\ext_{k}}{n_{k}}{\ext_{k+1}{n_{k}}} \IWishpdf{\ext_{k+1}}{\tilde{v}_{k+1}}{\tilde{V}_{k+1}} \diff\ext_{k+1} \\
	& \overset{\text{L\ref{lem:Approx_IW_as_W}}}{\approx} \IWishpdf{\ext_{k}}{v_{k|k}}{V_{k|k}}  \int \Wishpdf{\ext_{k}}{n_{k}-d-1}{\frac{\ext_{k+1}{n_{k}}}{(n-2d-2)(n-d-1)}} \IWishpdf{\ext_{k+1}}{\tilde{v}_{k+1}}{\tilde{V}_{k+1}} \diff\ext_{k+1} \\
	& \overset{\text{L\ref{lem:Integral_condW_IW}}}{=} \IWishpdf{\ext_{k}}{v_{k|k}}{V_{k|k}} \mathcal{GB}_{d}^{II}\left(\ext_{k}; \frac{n_k-d-1}{2},\frac{\tilde{v}_{k+1}-d-1}{2}, \frac{ n_{k} \tilde{V}_{k+1} }{(n-2d-2)(n-d-1)} ,\mathbf{0}_{d\times d}\right)  \\
	& \overset{\text{L\ref{lem:Approx_GB2_as_IW}}}{\approx} \IWishpdf{\ext_{k}}{v_{k|k}}{V_{k|k}} \IWishpdf{\ext_{k}}{\frac{\tilde{v}_{k+1}n_k -2(d+1)^2}{\tilde{v}_{k+1} + n_k -3d-3}}{ \frac{ n_{k} \tilde{V}_{k+1}}{\tilde{v}_{k+1} + n_k -3d-3} } \\
	& \overset{\text{L\ref{lem:ProductInverseWishart}}}{\propto}  \IWishpdf{\ext_{k}}{v_{k|k} + \frac{(v_{k+1|K}-v_{k+1|k}) -\frac{2(d+1)^2}{n_k}}{1 + \frac{v_{k+1|K}-v_{k+1|k} -3(d+1)}{n_k}}}{V_{k|k}+\frac{V_{k+1|K}-V_{k+1|k}}{1 + \frac{v_{k+1|K}-v_{k+1|k} -3d-3}{n_k}} }
\end{align}
\label{eq:ExtentSmoothing}
\end{subequations}
\rule{\textwidth}{1pt}
\end{figure*}

\subsection{Factorized model smoothing}
\label{app:FactSmoothing}

In the factorized case, there is no known analytical solution that gives a smoothed density of the desired factorized form; therefore approximations are necessary. Factorized density approximations are common in so called variational inference, and the factors are typically found by minimising the Kullback-Leibler divergence, see, e.g., \cite[Ch. 10]{Bishop:2006}. To find a factorised smoothed density, we apply Lemma~\ref{lem:KLminFactorisedApproximation} to the smoothed joint density $p(\xi_{k} | \setZ_{1:K})$, given in \eqref{eq:SmoothingBackwards}, and obtain the following two smoothing equations,
\begin{subequations}
\begin{align}
	p&\left(\sx_{k}|\setZ_{1:K}\right) = \int p(\xi_{k} | \setZ_{1:K}) \diff \ext_k   \\
	& = p\left(\sx_{k}|\setZ_{1:k}\right) \int \frac{p\left(\sx_{k+1}|\sx_{k}\right) p\left(\sx_{k+1}|\setZ_{1:K}\right)}{p\left(\sx_{k+1}|,\setZ_{1:k}\right) }  \diff \sx_{k+1} 
\end{align}
\label{eq:KLminimisingMarginalisationKinematicVector}
\end{subequations}
and
\begin{subequations}
\begin{align}
	p& \left(\ext_{k}|\setZ_{1:K}\right) = \int p(\xi_{k} | \setZ_{1:K}) \diff \sx_k  \\
	& = p\left(\ext_{k}|\setZ_{1:k}\right) \times   \\
	& \iint \frac{ p\left(\ext_{k+1}|\sx_{k},\ext_{k}\right) p\left(\ext_{k+1}|\setZ_{1:K}\right)}{p\left(\ext_{k+1}|\setZ_{1:k}\right) }  p\left(\sx_{k}|\setZ_{1:K}\right) \diff \ext_{k+1}  \diff \sx_k \nonumber
\end{align}
 \label{eq:KLminimisingMarginalisationExtentMatrix}
\end{subequations}
The proof of the marginalisation \eqref{eq:KLminimisingMarginalisationKinematicVector} is given in \eqref{eq:FeldmannKinematicVectorSmoothing}, and the proof of the marginalisation \eqref{eq:KLminimisingMarginalisationExtentMatrix} is given in \eqref{eq:FeldmannExtentMatrixSmoothing}.
\begin{figure*}[t]
\rule{\textwidth}{1pt}
\begin{subequations}
\begin{align}
	p&\left(\sx_{k}|\setZ_{1:K}\right) = \int p(\xi_{k} | \setZ_{1:K}) \diff \ext_k \\
	& = \int p(\xi_{k} | \setZ_{1:k}) \int \frac{p(\xi_{k+1}|\xi_{k}) p(\xi_{k+1}|\setZ_{1:K})}{p(\xi_{k+1}|\setZ_{1:k})} \diff \xi_{k+1}  \diff \ext_k \\
	& \overset{\text{A\ref{ass:IndependentKinematicTransition}}}{=} \int p\left(\sx_{k}|\setZ_{1:k}\right)p\left(\ext_{k}|\setZ_{1:k}\right) \iint \frac{p\left(\sx_{k+1}|\sx_{k}\right) p\left(\ext_{k+1}|\sx_{k},\ext_{k}\right) p\left(\sx_{k+1}|\setZ_{1:K}\right)p\left(\ext_{k+1}|\setZ_{1:K}\right)}{p\left(\sx_{k+1}|,\setZ_{1:k}\right)p\left(\ext_{k+1}|\setZ_{1:k}\right) } \diff \sx_{k+1}  \diff \ext_{k+1}  \diff \ext_k \\
	& = p\left(\sx_{k}|\setZ_{1:k}\right) \int \frac{p\left(\sx_{k+1}|\sx_{k}\right) p\left(\sx_{k+1}|\setZ_{1:K}\right)}{p\left(\sx_{k+1}|,\setZ_{1:k}\right) }  \diff \sx_{k+1} \int p\left(\ext_{k}|\setZ_{1:k}\right) \int \frac{ p\left(\ext_{k+1}|\sx_{k},\ext_{k}\right) p\left(\ext_{k+1}|\setZ_{1:K}\right)}{ p\left(\ext_{k+1}|\setZ_{1:k}\right) } \diff \ext_{k+1}  \diff \ext_k \\
	& = p\left(\sx_{k}|\setZ_{1:k}\right) \int \frac{p\left(\sx_{k+1}|\sx_{k}\right) p\left(\sx_{k+1}|\setZ_{1:K}\right)}{p\left(\sx_{k+1}|,\setZ_{1:k}\right) }  \diff \sx_{k+1}
\end{align}
\label{eq:FeldmannKinematicVectorSmoothing}
\end{subequations}
\rule{\textwidth}{1pt}
\end{figure*}
\begin{figure*}[t]
\rule{\textwidth}{1pt}
\begin{subequations}
\begin{align}
	p&\left(\ext_{k}|\setZ_{1:K}\right) = \int p(\xi_{k} | \setZ_{1:K}) \diff \sx_k \\
	& = \int p(\xi_{k} | \setZ_{1:k}) \int \frac{p(\xi_{k+1}|\xi_{k}) p(\xi_{k+1}|\setZ_{1:K})}{p(\xi_{k+1}|\setZ_{1:k})} \diff \xi_{k+1}  \diff \sx_k \\
	& \overset{\text{A\ref{ass:IndependentKinematicTransition}}}{=} \int p\left(\sx_{k}|\setZ_{1:k}\right)p\left(\ext_{k}|\setZ_{1:k}\right) \iint \frac{p\left(\sx_{k+1}|\sx_{k}\right) p\left(\ext_{k+1}|\sx_{k},\ext_{k}\right) p\left(\sx_{k+1}|\setZ_{1:K}\right)p\left(\ext_{k+1}|\setZ_{1:K}\right)}{p\left(\sx_{k+1}|,\setZ_{1:k}\right)p\left(\ext_{k+1}|\setZ_{1:k}\right) } \diff \sx_{k+1}  \diff \ext_{k+1}  \diff \sx_k \\
	& = p\left(\ext_{k}|\setZ_{1:k}\right) \iint \frac{ p\left(\ext_{k+1}|\sx_{k},\ext_{k}\right) p\left(\ext_{k+1}|\setZ_{1:K}\right)}{p\left(\ext_{k+1}|\setZ_{1:k}\right) } p\left(\sx_{k}|\setZ_{1:k}\right) \int \frac{p\left(\sx_{k+1}|\sx_{k}\right) p\left(\sx_{k+1}|\setZ_{1:K}\right) }{p\left(\sx_{k+1}|,\setZ_{1:k}\right) } \diff \sx_{k+1} \diff \ext_{k+1}  \diff \sx_k \\
	& = p\left(\ext_{k}|\setZ_{1:k}\right) \iint \frac{ p\left(\ext_{k+1}|\sx_{k},\ext_{k}\right) p\left(\ext_{k+1}|\setZ_{1:K}\right)}{p\left(\ext_{k+1}|\setZ_{1:k}\right) }  p\left(\sx_{k}|\setZ_{1:K}\right) \diff \ext_{k+1}  \diff \sx_k
\end{align}
\label{eq:FeldmannExtentMatrixSmoothing}
\end{subequations}
\rule{\textwidth}{1pt}
\end{figure*}

For the kinematic vector, we have that for Gaussian densities $p(\sx_{k+1}|\setZ_{1:K})$ and $p(\sx_{k+1}|\setZ_{1:k})$, see \eqref{eq:factorizedGIW}, and a Gaussian transition density $p(\sx_{k+1}|\sx_{k})$, see \eqref{eq:factTransitionDensity}, the smoothed kinematic state density is Gaussian with parameters given by the standard RTS-smoothing backwards step, given in, e.g., \cite[Thm. 8.2]{Sarkka:2013}. We get the result in Table~\ref{tab:factSmoothing}.

For the extent matrix, the smoothing \eqref{eq:KLminimisingMarginalisationExtentMatrix} does not have an analytical solution, and approximations are necessary. The derivation of the result in Table~\ref{tab:factSmoothing} is given in \eqref{eq:FeldmannExtentMatrixSmoothing2}.

\begin{figure*}[t]
\rule{\textwidth}{1pt}
\begin{subequations}
\begin{align}
	p&\left(\ext_{k}|\setZ_{1:K}\right) = p\left(\ext_{k}|\setZ_{1:k}\right) \iint \frac{ p\left(\ext_{k+1}|\sx_{k},\ext_{k}\right) p\left(\ext_{k+1}|\setZ_{1:K}\right)}{p\left(\ext_{k+1}|\setZ_{1:k}\right) }  p\left(\sx_{k}|\setZ_{1:K}\right) \diff \ext_{k+1}  \diff \sx_k \\
	& = p\left(\ext_{k}|\setZ_{1:k}\right) \iint \frac{ \Wishpdf{\ext_{k+1}}{n_{k}}{\frac{M(\sx_{k})\ext_{k} M^{\tp}(\sx_{k})}{n_{k}}} \IWishpdf{\ext_{k+1}}{v_{k+1|K}}{V_{k+1|K}} }{ \IWishpdf{\ext_{k+1}}{v_{k+1|k}}{V_{k+1|k}} }  \diff \ext_{k+1} p\left(\sx_{k}|\setZ_{1:K}\right) \diff \sx_k \\
	& \overset{\text{L\ref{lem:FractionInverseWishart}}}{\propto} p\left(\ext_{k}|\setZ_{1:k}\right)   \\
	& \quad \times \iint \Wishpdf{\ext_{k+1}}{n_{k}}{\frac{M(\sx_{k})\ext_{k} M^{\tp}(\sx_{k})}{n_{k}}} \IWishpdf{\ext_{k+1}}{v_{k+1|K}-v_{k+1|k}}{V_{k+1|K}-V_{k+1|k}}   \diff \ext_{k+1}  p\left(\sx_{k}|\setZ_{1:K}\right)  \diff \sx_k \nonumber\\
	& \overset{\text{L\ref{lem:WishartInverseWishart}}}{=} p\left(\ext_{k}|\setZ_{1:k}\right) \iint \IWishpdf{\ext_{k}}{n_{k}}{{n_{k}}{M^{-1}(\sx_{k})\ext_{k+1} M^{-T}(\sx_{k})}} \IWishpdf{\ext_{k+1}}{w}{W}   \diff \ext_{k+1}  p\left(\sx_{k}|\setZ_{1:K}\right) \diff \sx_k \\
	& \overset{\text{L\ref{lem:Approx_IW_as_W}}}{\approx} p\left(\ext_{k}|\setZ_{1:k}\right) \nonumber \\
	& \quad \times \iint \Wishpdf{\ext_{k}}{n_{k}-d-1}{ \frac{n_{k} M^{-1}(\sx_{k})\ext_{k+1} M^{-T}(\sx_{k})}{(n_k-d-1)(n_k-2d-2)}} \IWishpdf{\ext_{k+1}}{w}{W}   \diff \ext_{k+1}  p\left(\sx_{k}|\setZ_{1:K}\right) \diff \sx_k \\
	& \overset{\text{L\ref{lem:Integral_condW_IW}}}{=} p\left(\ext_{k}|\setZ_{1:k}\right) \int \mathcal{GB}_{d}^{II}\left(\ext_{k} ; \frac{n_k-d-1}{2}, \frac{w-d-1}{2},  \frac{n_{k} M^{-1}(\sx_{k})W M^{-T}(\sx_{k})}{(n_k-d-1)(n_k-2d-2)} , \mathbf{0}_{d\times d} \right)  p\left(\sx_{k}|\setZ_{1:K}\right) \diff \sx_k \\
	& \overset{\text{L\ref{lem:Approx_GB2_as_IW}}}{\approx} p\left(\ext_{k}|\setZ_{1:k}\right) \int \IWishpdf{\ext_{k}}{\frac{wn_k -2(d+1)^2}{w + n_k -3d-3}}{ \frac{ n_{k} M^{-1}(\sx_{k})W M^{-T}(\sx_{k})}{w + n_k -3d-3} }  \Npdfbig{\sx_{k}}{m_{k|K}}{P_{k|K}} \diff \sx_k \\
	& \overset{\text{\cite[Thm. 2]{GranstromO:2014}}}{\approx} p\left(\ext_{k}|\setZ_{1:k}\right) \int \IWishpdf{\ext_{k}}{\frac{wn_k -2(d+1)^2}{w + n_k -3d-3}}{ \frac{ n_{k} \mathbb{V}_{\sx_{k}} }{w + n_k -3d-3} } \Wishpdf{\mathbb{V}_{\sx_{k}}}{h}{h^{-1}C_4} \diff \mathbb{V}_{\sx_{k}}\\
	& \overset{\text{L\ref{lem:Integral_condIW_W}}}{\approx} p\left(\ext_{k}|\setZ_{1:k}\right) \mathcal{GB}_{d}^{II}\left( X_{k} ; \frac{h}{2}, \frac{1}{2} \frac{(w-d-1)(n_k-d-1)}{w + n_k -3d-3}, \frac{ n_{k} h^{-1}C_4}{w + n_k -3d-3} , \mathbf{0}\right) \\
	& \overset{\text{L\ref{lem:Approx_GB2_as_IW}}}{\approx} \IWishpdf{\ext_{k}}{v_{k|k}}{V_{k|k}} \IWishpdf{\ext_{k}}{\eta_{2}^{-1} \left(g -\frac{2(d+1)^2}{h+d+1}\right)}{\eta_{3}^{-1}C_{4} } \\
	& \overset{\text{L\ref{lem:ProductInverseWishart}}}{\propto} \IWishpdf{\ext_{k}}{v_{k|k} + \eta_{2}^{-1} \left(g -\frac{2(d+1)^2}{h+d+1}\right)}{V_{k|k} + \eta_{3}^{-1}C_{4} }
\end{align}


\label{eq:FeldmannExtentMatrixSmoothing2}
\end{subequations}
\rule{\textwidth}{1pt}
\end{figure*}

\bibliographystyle{IEEEtran}
\bibliography{PHD_ext_targ_track}
%
%
%

\end{document}